\documentclass{aastex}	
\usepackage{emulateapj5,onecolfloat} 

\newcommand {\xten}[1] {\mbox{$\times 10^{#1}$}}
\newcommand {\HST}     {{\em HST\/}}

\journalinfo{The Astronomical Journal, January 2002 ({\rm in press})}
\slugcomment{ }

\shortauthors{Parker et al.}
\shorttitle{HST Observations of Ceres}

\begin{document}

\twocolumn[

\title{Analysis of the First Disk-Resolved Images of Ceres from Ultraviolet
Observations with the Hubble Space Telescope}

\author{Joel~Wm.~Parker,\altaffilmark{1} S.~Alan~Stern}
\affil{Department of Space Studies, Southwest Research Institute,
Suite 426, 1050 Walnut Street, Boulder, CO 80302, USA}

\author{\medskip Peter C. Thomas}
\affil{Center for Radiophysics and Space Research, Cornell University, Ithaca,
NY 14853, USA}

\author{\medskip Michel~C.~Festou\altaffilmark{2}, William~J.~Merline, Elliot F. Young}
\affil{Department of Space Studies, Southwest Research Institute,
Suite 426, 1050 Walnut Street, Boulder, CO 80302, USA}

\author{\medskip Richard P. Binzel}
\affil{Department of Earth, Atmospheric, and Planetary Sciences, Massachusetts
Institute of Technology, Cambridge, MA 02139, USA}

\smallskip \and

\vspace*{-1ex} 

\author{Larry A. Lebofsky}
\affil{Lunar and Planetary Laboratory, University of Arizona, Space Sciences
Building, P.O. Box 210092 Tucson, AZ 85721-0092, USA}

\begin{abstract}

We present \HST\ Faint Object Camera observations of the asteroid 1~Ceres at
near-, mid-, and far-UV wavelengths ($\lambda = 3636$, 2795, and 1621~\AA,
respectively) obtained on 1995 June 25.  The disk of Ceres is well-resolved for
the first time, at a scale of $\sim 50$~km.  We report the detection of a
large, $\sim 250$~km diameter surface feature for which we propose the name
``Piazzi''; however it is presently uncertain if this feature is due to a
crater, albedo variegation, or other effect.  From limb fits to the images, we
obtain semi-major and semi-minor axes of $R_1=484.8 \pm 5.1$~km and $R_2=466.4
\pm 5.9$~km, respectively, for the illumination-corrected projected ellipsoid.
Although albedo features are seen, they do not allow for a definitive
determination of the rotation or pole positions of Ceres, particularly because
of the sparse sampling (two epochs) of the 9~hour rotation period.  From
full-disk integrated albedo measurements, we find that Ceres has a red spectral
slope from the mid- to near-UV, and a significant blue slope shortward of the
mid-UV.  In spite of the presence of Piazzi, we detect no significant global
differences in the integrated albedo as a function of rotational phase for the
two epochs of data we obtained.  From Minnaert surface fits to the near- and
mid-UV images, we find an unusually large Minnaert parameter of $k \approx
0.9$, suggesting a more Lambertian than lunar-like surface.

\end{abstract}


]

\altaffiltext{1}{{\tt joel@boulder.swri.edu}}

\altaffiltext{2}{Permanent address: Observatoire Midi-Pyr\'{e}n\'{e}es, 14
avenue E.  Belin, F31400 Toulouse, France}



\section{Introduction}

In the latter part of the 18th century, the Titius-Bode law led scientists to
believe there was a ``missing planet'' in the region between Mars and Jupiter.
On January 1st, 1801 --- the first day of the 19th century --- Giuseppe Piazzi
serendipitously discovered the first minor planet:  1~Ceres

Ceres is a G-type asteroid \citep{T84,BCCF87} that orbits the Sun with a 4.4~yr
period, a semi-major axis of 2.7~AU, and an eccentricity of 0.097.  It has a
rotational period of 9.075 hours \citep{LHZ89}.  \citet{L+86} reported that
thermal observations of Ceres before and after opposition indicate that it is a
prograde rotator (i.e., the pole is in the ecliptic North).  Ceres has an
absolute magnitude of $H = 3.32$, and a magnitude slope parameter $G = 0.12 \pm
0.02$ \citep{LM90}.  Ceres has measured optical colors of $U-B=0.43$ and
$B-V=0.72$, and a visual geometric albedo $\sim 0.10$ \citep{T89}.

\bigskip
\bigskip
\bigskip

In this paper we present {\em Hubble Space Telescope\/}/Faint Object Camera
(\HST/FOC) observations, which are the first images to resolve the disk of
Ceres sufficiently to detect surface features.  We present the observations, an
analysis of the photometric results, and we discuss a few tantalizing features
suggested by the data.  Though the images are not sufficient to provide
definitive new results regarding such issues as pole position and composition
(due to restricted temporal and wavelength coverage), they do provide a basis
data set to direct further observations that should answer these questions.


\bigskip

\section{Observations and Data Reduction \label{sec:obs}}

The observations we report here were made by \HST\ with the FOC over a 5-hour
period on 1995 June 25.  Table~\ref{tab:obs} describes the observations.

\begin{table*}
\caption{Summary of Observations \label{tab:obs}}
\begin{tabular}{llcclccc}
\hline \hline
\multicolumn{1}{c}{Data Set}  &
\multicolumn{1}{c}{Band} &
\multicolumn{1}{c}{$\lambda_{\rm center}$} &
\multicolumn{1}{c}{$\Delta \lambda$} &
\multicolumn{1}{c}{Filters} &
\multicolumn{1}{c}{Start Time} &
\multicolumn{1}{c}{Exp. Time} &
\multicolumn{1}{c}{Rotational\tablenotemark{a}\rule{0pt}{3ex}} \\
\multicolumn{1}{c}{ } &
\multicolumn{1}{c}{ } &
\multicolumn{1}{c}{(\AA)} &
\multicolumn{1}{c}{(\AA)} &
\multicolumn{1}{c}{ } &
\multicolumn{1}{c}{(1995 Jun 25, UT)} &
\multicolumn{1}{c}{(sec)} &
\multicolumn{1}{c}{Phase $\phi$ (deg)\rule[-1.5ex]{0pt}{0pt}} \\
\hline
x2og0101t & mid-UV  & 2795 & 134 & F4ND F275W F278M       & 10:02:56 & \ 716.5 & \ \ 0.00\rule{0pt}{3ex} \\
x2og0102t & near-UV & 3626 & 106 & F6ND F342W F1ND F372M & 10:22:47 & \ 910.5 & \  14.19 \\
x2og0103t & far-UV  & 1621 & 159 & F175W F152M           & 11:29:33 & \ 896.5 & \  58.26 \\
x2og0104t & far-UV  & 1621 & 159 & F175W F152M           & 11:49:28 &  1292.5 & \  73.61 \\
x2og0105t & mid-UV  & 2795 & 134 & F4ND F275W F278M       & 13:05:58 &  1016.5 &   122.67 \\
x2og0106t & near-UV & 3626 & 106 & F6ND F342W F1ND F372M & 13:30:49 &  1001.5 &   139.01 \\
x2og0107t & far-UV  & 1621 & 159 & F175W F152M           & 14:42:34 & \ 896.5 &   185.87 \\
x2og0108t & far-UV  & 1621 & 159 & F175W F152M           & 15:02:29 &  1292.5 &   201.22 \\
\hline \hline
\end{tabular}

\vspace*{-4ex}

\tablenotetext{a}{Rotational phase is relative to the first exposure and
assumes a rotational period of $P=9.075$~h \citep{LHZ89}.  The given values of
the rotational phase are measured at the midpoint of each exposure.}

\end{table*}

At the time of observations, the Sun-Ceres-Earth phase angle was
$\alpha=19.4$~deg, producing an observed illuminated fraction of 97.2\% of the
disk of Ceres; the corresponding defect of illumination was 0.012~arcseconds.

\newpage

Ceres' heliocentric distance was $r=2.57$~AU and its geocentric distance was
$\Delta=2.97$~AU.  At this distance, one arcsecond equates to about 2150~km, so
the 0.01435~arcsec width of one FOC pixel corresponded to a physical scale of
30.9~km on the surface of Ceres.  As stated in the \HST/FOC Instrument
Handbook, the FWHM of the FOC PSF for the filters we used in these observations
is approximately 0.03~arcsec, giving a Rayleigh criterion resolution of $\sim
0.024$~arcsec (the width of $\sim 1.7$~pixels) or a spatial resolution of $\sim
52$~km on the surface of Ceres.  From previous estimates of its size, we
expected Ceres to have an angular diameter of about 0.43~arcsec (the width of
30 FOC pixels), giving us over 700 FOC pixels covering the disk of Ceres.  This
is significantly better spatial resolution and sampling than has been
obtained previously with ground-based adaptive optics observations
\citep{SCM93, DFCH98}.

Our {\em HST\/}/FOC observations consisted of two sets of exposures in three
wavelength bands:  the far-UV (using a F175W+F152M filter combination;
$\lambda_c=1621$~\AA, $\Delta \lambda =159$~\AA), the mid-UV (F275W+F278M;
$\lambda_c=2795$~\AA, $\Delta \lambda =134$~\AA), and near-UV (F342W+F372M;
$\lambda_c=3626$~\AA, $\Delta \lambda =106$~\AA).  Neutral density filters were
used in the two longer-wavelength observations to keep the count rates at
reasonable levels.  We selected these UV bands because compared to visible
wavelengths they provide better resolution, ice diagnostics, and surface
contrast to search for features.

\begin{figure*}

\vspace*{8in}
\includegraphics{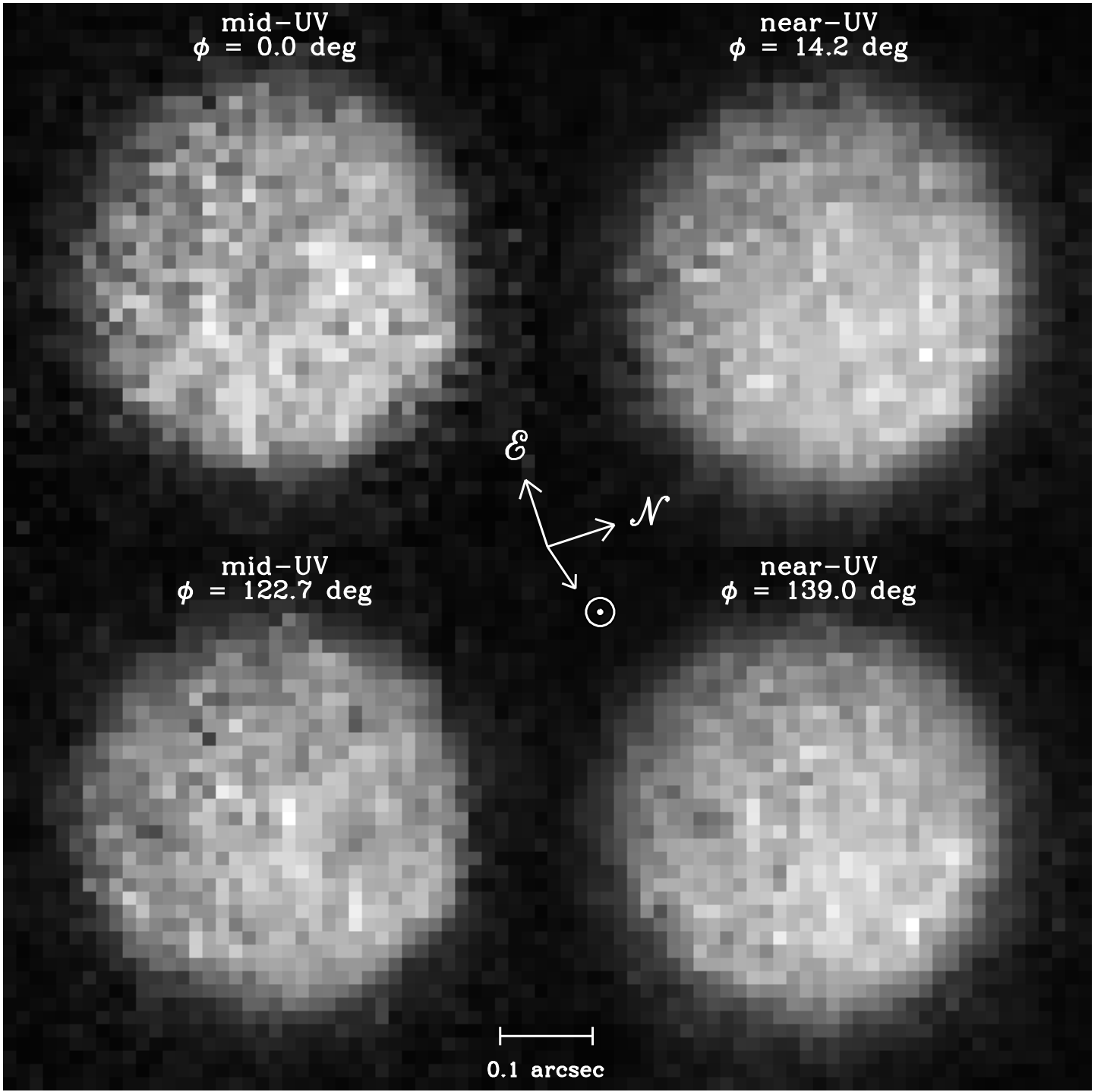}

\caption[fourims]{The mid-UV and the near-UV \HST/FOC images of
Ceres.  Vectors show the direction of the Sun and the image
orientation.  The values for the relative rotational phase, $\phi$, of
each image were calculated assuming a synodic rotational period of
$P=9.075$~h \citep{LHZ89}.  The images have been scaled such that they
are all normalized to their peak pixel value (given in
Table~\protect\ref{tab:phot}) for optimal contrast and ease of
cross-comparisons.  The Piazzi feature is near the center of the first
mid-UV image ($\phi = 0.0$~deg).
\label{fig:fourims} }

\end{figure*}

The FOC images, obtained with the COSTAR optical correction \citep{J+94}, were
reduced in the standard STScI Routine Science Data pipeline, which performed
flatfielding and geometric corrections.  The resulting images have a field of
view of $7.3 \times 7.3$~arcsec$^2$ ($512 \times 512$ pixels, with each pixel
$14.35 \times 14.35$ mas$^2$ in size).  Figure~\ref{fig:fourims} shows the four
mid-UV and near-UV FOC images of Ceres we obtained.  The far-UV images are not
shown here because they do not contain sufficient signal for displaying the
disk or detecting features.  However, as we show later, there is sufficient
signal in the far-UV images to measure total disk-integrated counts (and thus
determine the far-UV flux) from Ceres, which we compare to the total fluxes in
the other bands as a measure of Ceres' UV color slope.

\section{Discussion and Results \label{sec:disc}}

\subsection{Regarding Shape and Pole Position}

Ceres is the largest asteroid, but even after two hundred years of
observations, its size is still in dispute.  Estimates of the radius of
Ceres have ranged from 391~km \citep{B1900} to 600~km \citep{DKB74}.  {\em
Infrared Astronomical Satellite\/} ({\em IRAS\/}) observations made in 1983
imply a radius of $457 \pm 22$~km \citep{M86}.  More recent observations have
considerably improved measurements to quoted uncertainties better than 1\%, but
with values differing by at least a few percent.  The best direct size
measurements available are the stellar occultation observations of \citet{M+87}
and the adaptive optics images obtained by \citet{SCM93} and \citet{DFCH98}.
Millis et al. found projected equatorial and polar radii of $R_1=479.6 \pm
2.4$~km and $R_2=453.4 \pm 4.5$~km, respectively, and an equivalent radius of
$R=\sqrt{R_1 R_2} = 466.3 \pm 2.6$~km.  Saint-P\'{e} et al. find values that
imply $R_1 \sim 499$~km and $R_2 \sim 469$~km, giving $R = 484$~km.  The
Drummond et al.  observations were rotationally resolved, allowing them to
determine a fully triaxial shape with radii of $a=508$~km, $b=473$~km, and
$c=445$~km, with uncertainties of about 5~km; their projected ellipse values
are about 4\% larger than those determined from the occultation observations of
Millis et al..

We measured the disk center and the projected semi-major and semi-minor radii
of Ceres by performing ellipse fits to each of the four good-signal
\HST\ images; this method was previously used to obtain shape parameters of the
\HST/WFPC2 images of Vesta \citep{T+97}.  Each image was scanned along columns
or lines, depending on which is more nearly perpendicular to the limb.  The
resulting values were used to position a sharp edge along the scan; this was
fit using the predicted disk brightness and a Gaussian smear from disk
brightness to pixel brightness.  (The predicted brightness on the disk was
simply taken from the scan across the edge, where pixels that are 5--7 from the
initial edge are averaged for an ``on disk'' value.) The position of the sharp
edge was determined to 0.1~pixel-width by least squares matching of the
predicted brightness along the scan to the actual one as a function of the
modeled position of the sharp edge.  For these images, obtained at modest phase
angle, we performed a full-disk scan and corrected the down-sun (terminator)
position to a predicted limb position of the raw data, and then ellipses were
fit.  Fits of the half ellipses (illuminated limb only) produced similar
results, but have more uncertainty due to the greater uncertainty of the fit
center.  Table~\ref{tab:centers} gives the sub-Earth and sub-solar pixel
coordinates in the FOC images resulting from our fits.

\begin{table}[t]
\caption{Fitted Sub-Earth and Sub-Solar Pixel Coordinates
\label{tab:centers}}

\begin{center}
\vspace*{-2ex}
\begin{tabular}{lcc}
\hline \hline
\multicolumn{1}{c}{Data Set}  &
\multicolumn{1}{c}{Sub-Earth [$X_c,Y_c$]\tablenotemark{a}} &
\multicolumn{1}{c}{Sub-Solar [$X_s,Y_s$]\tablenotemark{b}} \\
\hline
x2og0101t & [197.80, 398.19] & [200.63, 393.95] \\
x2og0102t & [196.38, 396.50] & [199.21, 392.26] \\
x2og0105t & [192.41, 390.66] & [195.24, 386.42] \\
x2og0106t & [190.67, 389.53] & [193.50, 385.29] \\
\hline \hline
\end{tabular}
\end{center}

\vspace*{-6ex}

\tablenotetext{a}{The Center pixel of the disk of Ceres is determined using the
limb-fitting procedure described in the text.  The coordinate system has an
origin such that position [1,1] is the center of the lower-left pixel in the
FOC image.  The far-UV images have insufficient S/N to determine the disk
center.}

\tablenotetext{b}{The sub-solar position is $\Delta X = X_s - X_c = +2.83$,
$\Delta Y = -4.24$ pixel-widths from the sub-Earth coordinate.}

\end{table}

Using this method, we obtain average values for Ceres'
illumination-corrected, projected semi-major and -minor radii of
$R_1=484.8 \pm 5.1$~km and $R_2=466.4 \pm 5.9$~km, respectively, and
an equivalent radius of $R=475.5 \pm 3.9$~km.  Since at the time of
these observations the pole is within 1--18~deg of the plane of the
sky for all of the published pole solutions, these radii of the
projected ellipsoid are expected to be close to the true values.  Our
radii values are consistent with, though slightly larger than (by 1 to
$2\sigma$), the sizes obtained from the occultation measurements by
\citet{M+87}, and are in better agreement with the adaptive optics
values.  Some of the differences could be due to the effect of
differing rotational phase and sub-Earth latitude for the different
observations.  Observations taken at different epochs see the disk
cross-section at different sub-Earth latitudes and longitudes, so
comparisons are problematic since only the \citet{DFCH98} observations
provide an estimated 3-dimensional shape for Ceres.  However, for an
object rotating this slowly, the equilibrium shape is expected to be a
Roche ellipsoid\footnote{Note that this is in contradiction to the
results of \citet{DFCH98}, who find significantly differing values for
all three axes.  Their shape values would imply a lightcurve amplitude
a factor of two larger than that observed; they conjecture that albedo
variations could decrease the shape-induced lightcurve amplitude.}
with axes of $a = b > c$, so the maximum projected axis is equal to
the the equatorial axes; the minimum projected axis is between $a$ and
$c$.

Using our projected radii we can determine the mean density of Ceres for a
given mass.  There have been several mass estimates for Ceres based on its
perturbations on other asteroids \citep[e.g.,][]{S74,L88,G91,ST95,VR98,H99,M00}
and on Mars \citep{SH89}.  Table~1 shown by \citet{M00} provides a complete
list of mass estimates of Ceres, showing a range of masses from $6.7 \times
10^{-10} {\cal M}_\odot$ \citep{S70} to $4.3 \times 10^{-10} {\cal M}_\odot$
\citep{K96}, though most of the values are within the range 4.6--$5.0 \times
10^{-10} {\cal M}_\odot$.  \citet{M+87} estimate a mean density of $\rho = 2.7
\pm 0.1~{\rm gm~cm}^{-3}$ using one of the higher mass estimates of $(5.9 \pm
0.3) \times 10^{-10} {\cal M}_\odot$ \citep{S74}.  For this mass, we obtain a
similar density of $\rho = 2.6 \pm 0.2~{\rm gm~cm}^{-3}$.  Using a lower mass
estimate for Ceres of $(4.39 \pm 0.05) \times 10^{-10} {\cal M}_\odot$ from
\citet{H99}, our measurements imply a density $\rho = 1.90 \pm 0.05~{\rm
gm~cm}^{-3}$.

The implications of these different density values are significant.  Given the
measured shape (specifically, the observed difference between the projected
radii, $R_1 - R_2$), the higher density is consistent with the assumption that
Ceres is homogeneous and in hydrostatic equilibrium.  However, at the lower
density value, an object with the size and rotation period of Ceres would have
a difference in the semi-major and -minor axes of $a-c \gtrsim 40$~km due to
rotational flattening.  The difference of the \citet{M+87} projected radii is
$R_1 - R_2 = 26.2 \pm 5.1$~km, and the difference in our fit is even less:
$18.4 \pm 7.8$~km.  Therefore, our results appear more consistent with the
higher mass and density values: an object with the lower mean density value and
the size and rotation period of Ceres should have greater flattening than is
observed.  However, if the lower density value is correct, then the observed
lack of significant rotational flattening would imply that Ceres is not
internally homogeneous.  Another possible explanation for this discrepancy
between the predicted and measured flattening would be that the pole position
affected our measurement of the projected radii, such that the difference
between the true axes, $a-c$, is significantly larger than our observed
projected values, $R_1 - R_2$; if that were the case, then Ceres could be
homogeneous at the lower density.  However, as discussed below, such a pole
position is unlikely, again making our results more consistent with a higher
mass value.

\begin{figure*}
\vspace*{8in}
\includegraphics{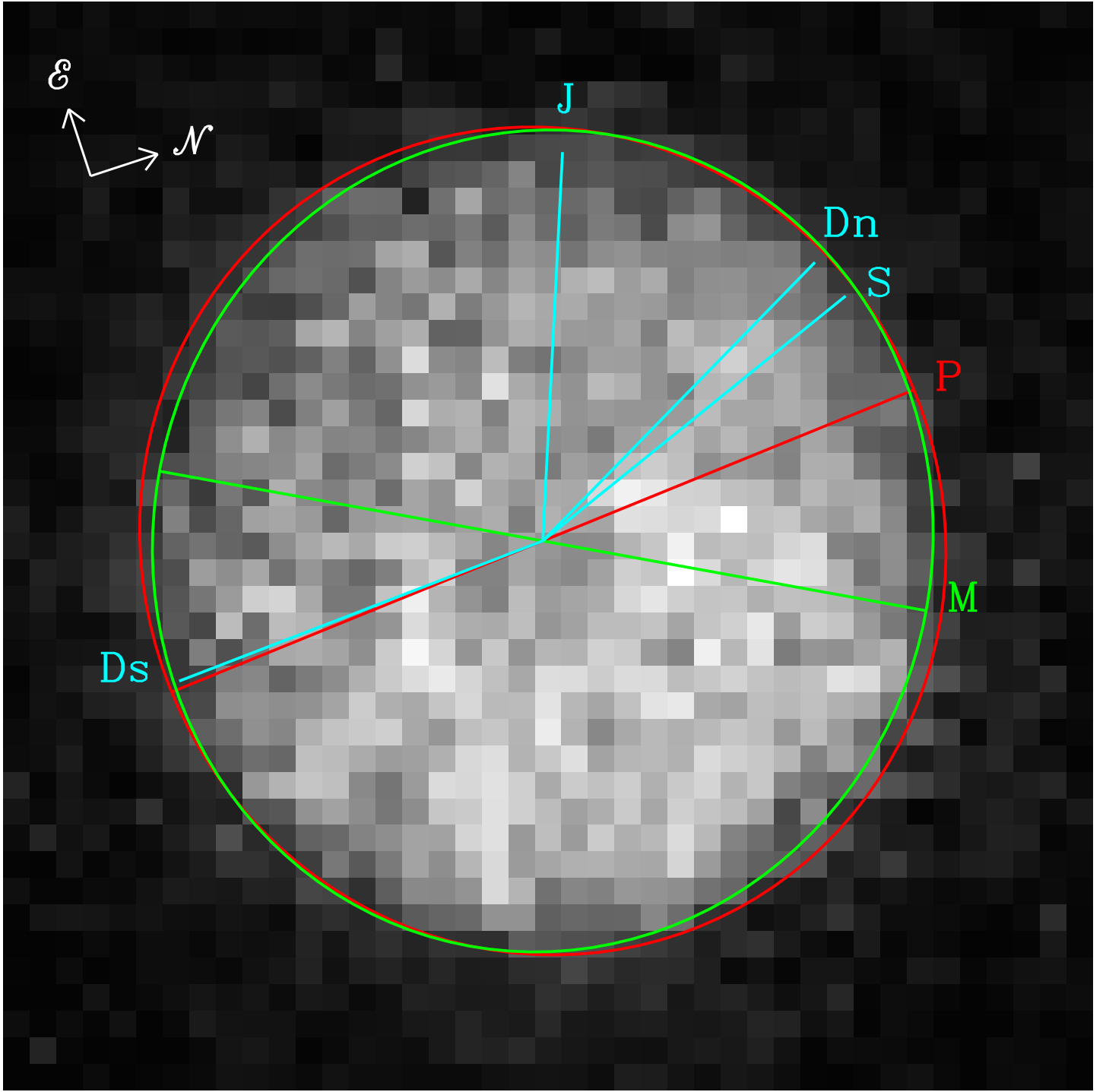}

\caption[oneim]{The first mid-UV image ($\phi = 0.0$~deg) from
Figure~\protect\ref{fig:fourims} showing the shape and pole position
angles determined from various sources: ``P'' = this paper, ``M'' =
\citet{M+87}, ``J'' = \citet{JKLR83}, ``S'' = \citet{SCM93}, and
``Dn'' and ``Ds'' are the \citet{DFCH98} north and south pole
positions, respectively.
\label{fig:oneim} }

\end{figure*}

Because of the sparse rotational sampling of our \HST\ data, we are
not able to unambiguously track sufficient surface features to obtain
a pole solution.  If we assume that the slight flattening we find in
the profile fit is due to rotation, then the best fit for our data put
the projected pole angle at about 4~deg from North, with an
uncertainty of roughly $\pm 15$~deg.  We show our pole position angle
and ellipse fit compared to those of other observations in
Figure~\ref{fig:oneim}.  These different values for the pole position
show a range of over 90~deg.  An initial analysis by \citet{Metal96}
of the motion of possible surface features seen in the FOC images
suggested that the pole position might be more consistent with the
\citet{JKLR83} than with the \citet{SCM93}, but that result could not
be convincingly re-established in our analysis.  The uncertainties in
our ellipsoidal fit do not allow combination with other results to
constrain the pole.  For example, the difference in projected axes
reported by of \citet{M+87} of $R_1 - R_2 = 26.2$~km could be reduced
to agree with our value of 18.4~km by viewing an ellipsoid of
$a=479.6, b=479.6, c=453.4$~km from 30~deg latitude.  However, the
published range of pole positions do not yield any viewing at greater
than 18~deg latitude.\footnote{For the instance of the \citet{SCM93}
solution, \citet{M+87} would have been observing from 3~deg latitude,
and this study would have been viewing at 10.9~deg.}  Thus, we may
only say the shape determined here, within our error bars, is
consistent with the occultation results, but does not further
constrain the spin pole.

\bigskip

\subsection{Regarding Surface Characteristics}

\subsubsection{The Piazzi Feature}

From examination of the images in Figure~\ref{fig:fourims}, no surface features
could be easily tracked because of the large rotational phase difference
between the pairs of FOC images within the same filter, and comparing images
between different filters can lead to misidentifications due to differing
reflectances at the different wavelengths.

However, while not trackable, there is one noticeable feature we are
compelled to discuss.  This is a significantly large, roughly
circular, and centrally darkened spot, which appears face-on in the
center of the first mid-UV image (rotational phase $\phi = 0.0$~deg;
upper left image in Figure~\ref{fig:fourims}), and covers about
40~pixels.  Using these same \HST/FOC images, this feature was
independently noted by \citet{LSWS98}.  Because of the low number of
counts, this is clearly not an artifact of nonlinear or wrapped count
effects.  It also is not a blemish from one of the FOC reseau marks,
which are $3 \times 3$~pixels in size (smaller than this feature); a
map of the reseau marks shows that none are closer than $\sim$27
pixel-widths from the center of the disk of the Ceres image.

\begin{figure}[t]

\plotone{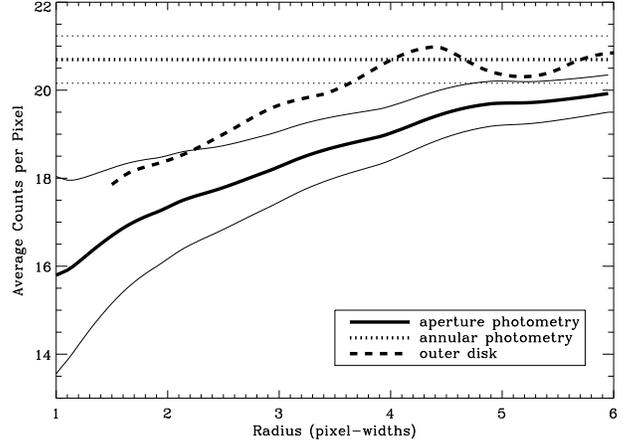}

\caption[phot_profile]{The photometric profile of the ``Piazzi''
feature on the first mid-UV image ($\phi = 0.0$~deg) from
Figure~\protect\ref{fig:fourims}.  The solid lines show the average
number of counts per pixel (thick line) and associated photometric
uncertainty (thin lines) within a circular aperture as a function of
aperture radius.  Similarly, the dashed line shows the average number
of counts within a 1-pixel-wide photometric annulus.  The horizontal
dotted lines show the average number of counts and associated errors
in a region beyond the feature.  Note the annular photometry of the
feature matches this background value at a radius of
$\sim$4~pixel-widths, which is our estimate for the mean radius of the
feature.
\label{fig:phot_profile} }

\end{figure}

To determine the reality of this feature, we compared the average
number of counts per pixel within the feature (allowing the radius and
center point to vary) to the average number of counts per pixel in an
annulus outside of the measurement aperture.  We also repeated the
same measurements for a annular ring rather than a photometric
aperture.  The results of these measurements are shown in
Figure~\ref{fig:phot_profile} and give us the following parameters:
{\em (i)\/} the center point has an albedo about 30\% lower than in
the region surrounding the feature, and it monotonically increases to
the edge of the feature; {\em (ii)\/} the optimal diameter of the
feature is 8~pixel-widths (consistent with our by-eye estimate of
7~pixel-widths), or about 250~km; and {\em (iii)\/} the center of the
dark feature is roughly 0.5 and $-0.5$ pixel-widths in the X and Y
directions, respectively, from the center of the disk image as
determined from our limb fits described earlier.

We also used the Kolmogorov-Smirnov (K-S) test \citep{PTVF92} to
examine the statistical significance of this feature.  We generated a
featureless model of Ceres based on the Minnaert surface fit discussed
in Section~\ref{sec:Minnaert} and added Poisson noise to that model.
We generated a cumulative distribution function based on 1000 such
random models, which was then used to compare to our data as well as
to 1000 more individual noise models to determine the distribution of
K-S statistics.  Comparing the FOC image to the models indicates a
significant ($\sim 6 \sigma$) difference, implying that this feature
is not likely due to random noise.  To check that this difference is
not just due to a poor model, we repeated the same tests after masking
out the feature's pixels in both the FOC image and in the models; in
this case the data agreed well ($\sim 2 \sigma$) with the models,
indicating that the model is a good match to the image, and only the
pixels in the area of this feature are statistically deviant.  To test
that this change was not just an effect of masking out pixels from the
analysis, we masked out other portions of the disk, but in those
cases, the difference between the data and model remained large
($\sim$ 5--9$\sigma$), again indicating that only the region
containing the feature is statistically unique.  As a final check, we
ran all the same tests on the second mid-UV image (labeled ``$\phi =
122.7$~deg'' in Figure~\ref{fig:fourims}).  Since that image does not
show any obvious feature, we would expect that the model would fit the
data in all cases (whether we were using all pixels, or masking out
any region) better than the 5--9$\sigma$ deviations found for the
first image with the feature not masked out.  We found that the data
in the second image were indeed better fit to the model in all cases,
typically at the 3--4$\sigma$ level (only one case had a $5\sigma$
deviation), but not as good a match as one might hope.  This could
indicate that the model does not fit the data for the second mid-UV
image quite as well, or perhaps the feature had rotated to another
area of the visible disk and, though not easily identifiable by eye,
is still statistically significant.  However, these tests indicate
that the models fit reasonably well.

Although our analysis cannot determine the nature of this feature,
i.e., whether it is a crater, an albedo spot, or something else, we
believe its existence is sufficiently established (to be
definitive, we need higher-resolution images with adequate time
sampling to see the motion of surface features).  With that in mind,
we propose the name ``Piazzi'' for this feature, in honor of the
discoverer of Ceres.

\bigskip

\subsubsection{The Geometric Albedo}

\begin{table*}[ht]
\caption{Summary of Photometry \label{tab:phot}}
\tabletypesize{\small}

\begin{center}
\vspace*{-2ex}
\begin{tabular}{llcccccc}
\hline \hline
\multicolumn{1}{c}{Data Set} &
\multicolumn{1}{c}{Band} &
\multicolumn{1}{c}{Exp. Time} &
\multicolumn{1}{c}{PHOTFLAM\tablenotemark{a}} &
\multicolumn{2}{c}{Counts} &
\multicolumn{1}{c}{Flux\tablenotemark{c}} &
\multicolumn{1}{c}{Geometric Albedo\tablenotemark{d}\rule{0pt}{3ex}} \\
\cline{5-6}
\multicolumn{1}{c}{} &
\multicolumn{1}{c}{} &
\multicolumn{1}{c}{(sec)} &
\multicolumn{1}{c}{} &
\multicolumn{1}{c}{Peak} &
\multicolumn{1}{c}{Total\tablenotemark{b}} &
\multicolumn{1}{c}{(erg s$^{-1}$ cm$^{-2}$ \AA$^{-1}$)} &
\multicolumn{1}{c}{\rule[-1.5ex]{0pt}{0pt}} \\
\hline
x2og0102t & near-UV & \ 910.5   & 9.720899\xten{-15} &  79.1 &   37793 & 4.035\xten{-13} &
 $0.0558 \pm 0.0010$ \\
x2og0106t & near-UV &  1001.5   & 9.720899\xten{-15} &  86.8 &   41088 & 3.988\xten{-13} &
 $0.0552 \pm 0.0010$ \\
x2og0101t & mid-UV  & \ 716.5   & 3.302684\xten{-15} &  28.6 &   13152 & 6.062\xten{-14} &
 $0.0298 \pm 0.0006$ \\
x2og0105t & mid-UV  &  1016.5   & 3.302684\xten{-15} &  42.5 &   18068 & 5.871\xten{-14} &
 $0.0288 \pm 0.0006$ \\
x2og0103t & far-UV  & \ 896.5   & 6.379518\xten{-15} & \ 5.8 & \ \ 677 & 4.820\xten{-15} &
 $0.1238 \pm 0.0128$ \\
x2og0104t & far-UV  &  1292.5   & 6.379518\xten{-15} & \ 4.9 & \ \ 468 & 2.310\xten{-15} &
 $0.0593 \pm 0.0068$ \\
x2og0107t & far-UV  & \ 896.5   & 6.379518\xten{-15} & \ 4.8 & \ \ 575 & 4.089\xten{-15} &
 $0.1059 \pm 0.0091$ \\
x2og0108t & far-UV  &  1292.5   & 6.379518\xten{-15} & \ 5.0 & \ \ 560 & 2.764\xten{-15} &
 $0.0710 \pm 0.0069$ \\
\hline \hline

\end{tabular}
\end{center}

\vspace*{-6ex}

\tablenotetext{a}{ PHOTFLAM is the inverse sensitivity factor used to
convert from units of (counts~sec$^{-1}$) to units of (erg s$^{-1}$
cm$^{-2}$ \AA$^{-1}$) as described in the HST Data Handbook.  The
values listed here are the calibrated factors obtained from the
processed image headers.}

\tablenotetext{b}{ Total Counts are sky-subtracted, disk-integrated
totals within a 0.29~arcsec (20~pixel-width) radius aperture.}

\tablenotetext{c}{ Flux = (PHOTFLAM) $\times$ (Total Counts) /
(Exp. Time), and is the $F_{\rm obj}(\lambda)$ term in
Equation~\ref{eq:albedo}.}

\tablenotetext{d}{ The albedo values are calculated using our
equivalent radius of $R=475.5$~km.  Quoted uncertainties represent
$1\sigma$ values were determined from Poisson statistics-based
aperture photometry values, the uncertainty in the measured equivalent
radius, and sensitivity to how the sky background id determined (which
is particularly significant for the very faint far-UV images).}

\end{table*}

The albedo of Ceres was determined in the standard manner and our results
calculated from the FOC data are given in Table~\ref{tab:phot}.  The geometric
albedo is defined as:
\begin{equation}
\label{eq:albedo}
p(\lambda,\alpha) = \frac{r^{2} \Delta^{2}}{R^{2}_{\rm obj}}
                      \left[\frac{F_{\rm obj}(\lambda)}
                                 {F_{\sun}(\lambda)}\right]
                      f_{\alpha}
\end{equation}

\noindent
where $r$ and $\Delta$ are the heliocentric (in AU) and geocentric (in km)
distances of the object at the time of the observation (these values for Ceres
are given in Table~\ref{tab:phot}), $R_{\rm obj}$ is the object's radius in km
(we use on our measured equivalent radius of $R=475.5$~km), $F_{\rm
obj}(\lambda)$ is the measured flux, and $F_{\sun}(\lambda)$ is the solar
flux at 1~AU over the same bandpass.  For $F_{\rm obj}$ we use the
background-subtracted disk-integrated flux calculated from the aperture
photometry measured total counts listed in Table~\ref{tab:phot}.  The
appropriate UV solar flux for these observations was obtained from the UARS
satellite UV spectrometer, and the spectrum was modified by the FOC instrument
response function using the {\sc synphot} tasks in {\sc iraf}.  The parameter
$f_{\alpha}$ is the dimensionless correction for solar phase angle.  The UV
phase function for Ceres is unknown; for our analysis here, we will use the
value of the visible phase function: $f_{\alpha}=2.56$, which was calculated
using the equations of \citet{Betal89} with a phase angle of $\alpha=19.4$~deg
and slope parameter of $G=0.12$ \citep{LM90}.

The albedo uncertainties quoted in Table~\ref{tab:phot} are $1\sigma$-values
based on the uncertainties in the equivalent radius of Ceres and in the
aperture photometry using a 0.29~arcsec (20~pixel-widths) radius aperture that
has been sky-subtracted using the modal value in a sky annulus with radii of
0.43--0.72~arcsec (30--50~pixel-widths);\footnote{Because the signal in the
far-UV images was insufficient for limb-fitting, we were not able to determine
the center coordinates of the disk in those images.  Therefore, in our
calculations of the disk-integrated flux, we used the center coordinates from
the near-UV images; the 0.29~arcsec radius photometry aperture allowed for any
positional shift (i.e., note that the shift between contiguous mid-UV and
near-UV images is about 2~pixels-widths, less than 0.03~arcsec, and the radius
of Ceres is about 0.22~arcsec).  Also, because of the very low signal-to-noise
in the far-UV images, the measured flux is {\em extremely\/} sensitive to the
estimated sky value, so the uncertainties quoted here also include the measured
scatter of values using sky annuli of different sizes.}  the FOC absolute
photometric error is roughly 10--15\%.

\begin{figure}[hb]

\plotone{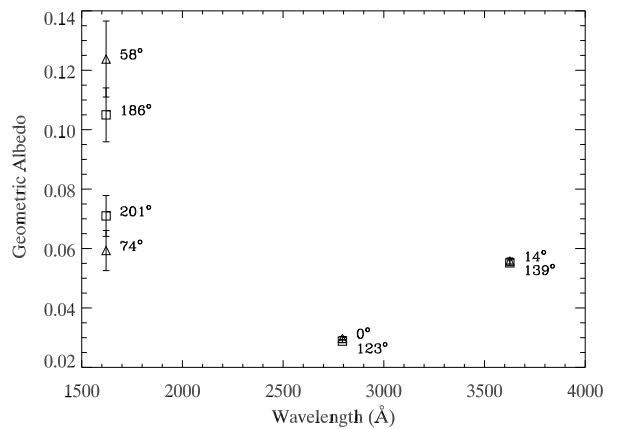}

\caption[albedo]{The UV geometric albedo of Ceres determined
from the \HST\ observations using a phase correction factor of
$f_{\alpha} = 2.56$ in Equation~\protect~\ref{eq:albedo}.  The
triangles ($\triangle$) indicate data from the first epoch of
observations, and the squares ($\square$) indicate data from the
second epoch.  The number by each point indicates the rotational
phase, $\phi$, of that observation.  $1\sigma$ error bars are plotted
representing aperture photometry and equivalent radius uncertainties
only; for the near-UV and mid-UV data the error bars are comparable to
the size of the plotted symbols.
\label{fig:albedo} }

\end{figure}

\bigskip
\bigskip

The albedo values and their uncertainties are plotted in
Figure~\ref{fig:albedo}.  Even within the lower limit uncertainties shown, the
albedo at the mid- and near-UV wavelengths display no evidence of rotational
phase dependent variation.  However, the data do show that Ceres has a red
color slope longward of the mid-UV (0.032 per 1000~\AA\ for $\lambda \gtrsim
2800$~\AA); Ceres has a red color slope (though not as steep) through the
visible bands as well \citep{BM85}.  Even within the large uncertainties in the
far-UV photometry, there is a significant blue slope ($-0.051$ per
1000~\AA\ for $\lambda \lesssim 2800$~\AA) from the far-UV to the mid-UV,
indicative of a minimum around 2800~\AA.  This break in slope is consistent
with the differing mechanisms for formation of features as discussed by
\citet{GBC89}: absorption shortward of $\sim 2500$~\AA\ results from
electron-hole pair formation, whereas the longer wavelength ($\sim 2500$ to
3500~\AA) absorptions are produced primarily by charge-transfer mechanisms.
There is no sign of significant absorption due to water across the disk, as
that feature would occur in the far-UV filter, and no significant drop is seen
in that bandpass.  However, this does not rule out the possibility of water ice
at the poles.

\citet{BM85} analyze {\em International Ultraviolet Explorer\/} ({\em IUE\/})
spectra of Ceres combined with data at longer wavelengths to determine albedo
values from 2100--10000~\AA.  Their values of $p \approx 0.05$ at near-UV
wavelengths and $p \approx 0.03$ at mid-UV wavelengths are in excellent
agreement with our results.  \citet{RB94} also provide albedo measurements
of Ceres from {\em IUE\/} observations in the wavelength range 2450--3150~\AA,
which nicely brackets our mid-UV measurements.  Their calculated albedo values
range from $p=0.024$--0.030, with an interpolated value of $p=0.025 \pm 0.001$
at our mid-UV wavelength (2795~\AA) whereas our averaged value is $p=0.029 \pm
0.001$.  The small difference could be due to values used for the phase angle
correction (as noted above, we use the visible phase function correction for
these UV data).  The agreement of these {\em IUE\/} values and our near-UV and
mid-UV results give us additional confidence in the reality of the large blue
slope we find in the far-UV. \citet{RB94} point out that G-type asteroids have
higher albedos than C-type asteroids in the visible, but the UV albedos of the
two type are similar.  However, our \HST\ observations also provide the first
published far-UV data of any asteroid, so we cannot compare the far-UV blue
slope to any other asteroid observations in order to understand how unique is
this feature.

\bigskip

\subsubsection{The Minnaert Parameters \label{sec:Minnaert}}

Since the FOC images typically resolve Ceres' disk into over 700 sunlit pixels,
we can calculate the incidence and emission angles for each pixel, and in
principle determine the Minnaert parameter, $k$, for Ceres.  In practice, the
brightness measured at each pixel will be a combination of Ceres' intrinsic
albedo distribution as well as the surface's scattering properties.  We
therefore compare full-disk solutions for $k$ for each of the near- and mid-UV
images, which should give us results that are reasonably insensitive to surface
albedo variations given the large number of illuminated pixels and having
images of Ceres at different rotational phases.

The Minnaert law \citep{M41} is an empirical approximation describing
scattering from a surface.  It has the form:

\begin{equation}
\label{eq:minnaert}
I/F = B_0  \mu_0^k  \mu^{k-1}, 
\end{equation}

\noindent
where $\mu_0$ and $\mu$ are cosines of the angles of emission and incidence,
respectively, and $B_0$ and $k$ are the two Minnaert parameters (central $I/F$
and limb darkening parameter, respectively).  The parameter $k$ typically
depends strongly on the phase angle, although for a Lambert surface $k=1$ for
all phase angles.  When fitting the model to the data, there is a correlation
between $k$ and the estimated radius as discussed below; we use our effective
radius value of $R=475.5$~km (equivalent to 15.4 pixel-widths), although the
fact that Ceres' disk is not circular contributes to the uncertainty in the
result.

We performed fits to the data using two independent methods: the $\chi^2$
minimization used by \citet{VHST89} in their analysis of Voyager~2 observations
of Uranian satellites, and fitting a linearized logarithmic version of
Eq.~\ref{eq:minnaert}.\footnote{\citet{VHST89} warn against linearizing
Eq.~\ref{eq:minnaert} by taking its logarithm, pointing out that this method
does not preserve the proper weighting of the data points.  However, one can
preserve the proper weighting by appropriate propagation of the errors;
specifically, if a pixel has a value of $N$ and an error of $\sigma$, then its
error in the log version of the Minnaert law is simply $\sigma/N$.}  Both
methods agree within the estimated 2$\sigma$ uncertainties;
Table~\ref{tab:minnaert} shows the results of our fits.  If we vary the radius
by $\pm 10$~km, the fitted values of $k$ typically vary by $\pm 0.04$, which
are on the order of the uncertainties of the fits listed in
Table~\ref{tab:minnaert}.  The values for $B_0$ do not appear to be correlated
with radius.

\begin{table}[t]
\caption{Fit Results for Minnaert Coefficients \label{tab:minnaert}}
\begin{center}
\vspace*{-2ex}
\begin{tabular}{llrr}
\hline
\multicolumn{1}{c}{Data Set} &
\multicolumn{1}{c}{Band} &
\multicolumn{1}{c}{$B_0$} &
\multicolumn{1}{c}{$k$} \\
\hline \hline
x2og0101t &  mid-UV & $22.8 \pm 0.9$ & $0.89 \pm 0.03$ \\
x2og0102t & near-UV & $65.4 \pm 2.5$ & $0.90 \pm 0.03$ \\
x2og0105t &  mid-UV & $32.1 \pm 1.3$ & $0.92 \pm 0.04$ \\
x2og0106t & near-UV & $71.4 \pm 2.7$ & $0.92 \pm 0.03$ \\
\hline \hline
\end{tabular}
\end{center}

\vspace*{-8ex}

\tablecomments{The correlation coefficient between $B_0$ and $k$ was 0.7. \\
Assumed radius is $R=475.5$~km = 0.22~arcsec.}

\end{table}

The limb darkening parameter, $k$, is typically a strong function of the phase
angle at the instant the observations were made, e.g., see Fig.~4 of
\citet{VHST89}. Ceres' phase angle was $\alpha=19.4$~deg in these
observations.  Regardless of the phase angle, however, Ceres' limb parameter is
quite high (around $k=0.9$) relative to $k$ values for Uranian, Galilean, and
Saturnian satellites \citep{B84, VHST89}.  The simplest interpretation is that
Ceres scatters more like a Lambertian surface ($k = 1$) than a lunar-like
surface ($k = 0.5$), which implies that much of the light we see from the
surface of Ceres undergoes multiple reflections.  Interestingly, \citet{SCM93}
find just the opposite in their infrared images: they find that the
center-to-limb brightness profile  in the $H$ and $K$ bands is better fit by a
(lunar-like) flat disk than a Lambertian sphere.  This disagreement could be
the result of their lower resolution and/or possible surface reflectance
differences between the IR and the UV wavelengths.  \citet{HM97} also find a
lower value for the Minnaert parameter, $k=0.61$, using a technique of
analyzing the signal modulation of {\em Hipparcos Satellite} data (Ceres is not
resolved in their data).  Although they use an average phase angle of
$\alpha=18$~deg, a value similar to ours, they use the smaller {\em
IRAS\/}-based radius, $R=456.5$~km, which would account for much of the
difference (their $\Delta k / \Delta R \approx 0.092$~km$^{-1}$): using the
radii values we find from our \HST\ data, the calculations of \citet{HM97}
would yield values of $k=0.74-0.87 \pm 0.05$, in better agreement with our
results.


\clearpage

\section{Conclusions}

Our analysis of \HST/FOC near-, mid-, and far-UV images of Ceres have produced
the following results:
\begin{enumerate}

\item These are the first, well-resolved ($\sim 50$~km) images of Ceres.

\item We have made a detection of an apparently large, $\sim 250$~km diameter
surface feature for which we propose the name ``Piazzi''.

\item The semi-major and semi-minor axes are $R_1=484.8 \pm 5.1$~km and
$R_2=466.4 \pm 5.9$~km, respectively, for the projected ellipsoid.

\item Disk-integrated photometry in the three bandpasses show that Ceres has
averaged geometric albedo values of:  $p=0.056$ in the near-UV, $p=0.029$ in
the mid-UV, and $p=0.090$ in the far-UV.  These values give a red spectral
slope (0.032 per 1000~\AA) from the mid- to near-UV, and a significant a blue
slope ($-0.052$ per 1000~\AA) shortward of the mid-UV.  The relatively
high albedo value in the far-UV implies there is no significant
absorption due to ice across the disk.

\item We detect no significant global differences in the full-disk integrated
albedo as a function of rotational phase for the two epochs of data we
obtained.

\item From Minnaert surface fits to the near- and mid-UV images, we find an
unusually large Minnaert parameter of $k \approx 0.9$, suggesting a more
Lambertian than lunar-like surface.

\end{enumerate}

It is clear that more such imaging observations are needed with better sampling
of the rotation period to finally resolve the continuing, long-standing
uncertainty in Ceres' pole position, and to obtain more information on the
intriguing Piazzi feature detected in these observations.  Also, more sensitive
observations in the far-UV should be obtained to search for surface ice on
Ceres and map its distribution; the far-UV bandpass is well-matched to the
strong water ice absorption near 1650~\AA\ as seen both in laboratory data
\citep[e.g.][]{H71} and in the rings of Saturn \citep{WC88}.  Such observations
could be obtained with the next generation of HST imaging space instruments,
allowing us to probe ever more deeply into the nature of this largest and
longest known asteroid.

\bigskip

\acknowledgements

Thanks to Leslie~Young and Dan~Durda for providing comments on early drafts of
this paper, Hal~Levison for several useful discussions, Ted~Bowell for
additional feedback, and an anonymous referee for helpful suggestions.  This
work is based on observations with the National Aeronautics and Space
Administration/European Space Agency {\em Hubble Space Telescope\/} obtained at
the Space Telescope Science Institute, which is operated by the Association of
Universities for Research in Astronomy, Incorporated, under NASA Contract
NAS5-26555.  Support for this work was provided by NASA through STScI Grant
GO-5842.



\begin{thebibliography}{dummy}

\bibitem[Altenhoff et al.(1996)]{ABSSV96} Altenhoff, W. J., Baars, J. W. M.,
	Schraml, J. B., Stumpff, P., \& Von Kap-Herr, A. 1996, \aap, 309, 953

\bibitem[Barnard(1900)]{B1900} Barnard, E. E. 1900, \mnras, 60, 261


\bibitem[Barucci et al.(1987)]{BCCF87} Barucci, M.\ A., Capria, M.\ T., 
Coradini, A., \& Fulchignoni, M.\ 1987, Icarus, 72, 304 


\bibitem[Bowell et al.(1989)]{Betal89} Bowell, E., Hapke, B., Domingue, D.,
Lumme, K., Peltoniemi, J., \& Harris, A. W.\ 1989, in Asteroids II, eds. R. P.
Binzel, T.  Gehrels, and M. S. Matthews (University of Arizona Press, Tucson),
524

\bibitem[Buratti(1984)]{B84} Buratti, B.\ J.\ 1984, Icarus, 59, 392

\bibitem[Butterworth \& Meadows(1985)]{BM85} Butterworth, 
P.~S.~\& Meadows, A.~J.\ 1985, Icarus, 62, 305 


\bibitem[Drummond et al.(1998)]{DFCH98} Drummond, J. D., Fugate, R. Q.,
	Christou, J. C., \& Hege, E.  K.  1998, Icarus, 132, 80

\bibitem[Dunham et al.(1974)]{DKB74} Dunham, D. W., Killen, S. W., \& Boone,
	T.  L. 1974 \baas, 6, 432

\bibitem[Gaffey et al.(1989)]{GBC89} Gaffey, M.\ J., Bell,
J.\ F., \& Cruikshank, D.\ P.\ 1989, in Asteroids II, eds. R. P. Binzel, T.
Gehrels, and M. S. Matthews (University of Arizona Press, Tucson), p. 98

\bibitem[Goffin(1991)]{G91} Goffin, E.\ 1991, \aap, 249, 563

\bibitem[Hestroffer \& Mignard(1997)]{HM97} Hestroffer, D.~\& Mignard,
F.\ 1997, ESA SP-402: Hipparcos - Venice '97, 402, 173

\bibitem[Hilton(1999)]{H99} Hilton, J.~L.\ 1999, \aj, 117, 1077

\bibitem[Hudson(1971)]{H71} Hudson, R.D., 1971, Rev.~Gephys. \& Space Phys., 9,
	360

\bibitem[Jedrzejewski et al.(1994)]{J+94} Jedrzejewski, R. I., Hartig, G.,
	Jakobsen, P., Crocker, J. H., \& Ford, H. C. 1994 \apj, 435, L7

\bibitem[Johnston et al.(1982)]{JSW82} Johnston, K. J., Seidelmann, P. K., \&
	Wade, C. M. 1982, \aj, 87, 1593

\bibitem[Johnson et al.(1983)]{JKLR83} Johnson, P. E., Kemp, J. C., Lebofsky,
	M. J., \& Rieke, G.  H.  1983, Icarus, 56, 381

\bibitem[Kuzmanoski(1996)]{K96} Kuzmanoski, M.\ 1996, IAU Symp.~172: Dynamics,
Ephemerides, and Astrometry of the Solar System, 172, 207

\bibitem[Lagerkvist et al.(1989)]{LHZ89} Lagerkvist, C.-I.,
Harris, A.\ W., \& Zappala, V.\ 1989, in Asteroids II, eds. R. P.  Binzel, T.
Gehrels, and M. S. Matthews (University of Arizona Press, Tucson), 1162

\bibitem[Lagerkvist \& Magnusson(1990)]{LM90} Lagerkvist, C.-I. \& Magnusson,
	P. 1990, A\&AS, 86, 119

\bibitem[Landgraf(1988)]{L88} Landgraf, W.\ 1988, \aap, 191, 161

\bibitem[Landis et al.(1998)]{LSWS98} Landis, R. R., Stern, S. A., Wood, C. A.
\& Storrs, A. D. 1998, Lunar and Planetary Science Conference, 29, 1937

\bibitem[Lebofsky et al.(1986)]{L+86} Lebofsky, L. A., et al. 1986, Icarus, 68,
239

\bibitem[Matson(1986)]{M86}  Matson, D. L. (ed.) 1986, JPL Internal Document
Number D-3698

\bibitem[Merline et al.(1996)]{Metal96} Merline, W. J., Stern, S. A., Binzel,
R. P., Festou, M. C., Flynn, B. C. \& Lebofsky, L. A.  1996, \baas, 28, 1101

\bibitem[Michalak(2000)]{M00} Michalak, G.\ 2000, \aap, 360, 363

\bibitem[Millis et al.(1987)]{M+87} Millis, R. L., et al. 1987, Icarus, 72, 507

\bibitem[Minnaert(1941)]{M41} Minnaert, M.\ 1941, \apj, 93, 403

\bibitem[Press et al.(1992)]{PTVF92} Press, W. H., Teukolsky, S. A.,
	Vetterling, W. T., \& Flannery, B. P. 1992, Numerical Recipes in
	Fortran:  The Art of Scientific Computing

\bibitem[Roettger \& Buratti(1994)]{RB94} Roettger, E.E., \& Buratti, B.J.
1994, Icarus, 112, 496

\bibitem[Saint-P\'{e} et al.(1993)]{SCM93} Saint-P\'{e}, O., Combes, M., \&
	Rigaut, F.  1993 Icarus, 105, 271

\bibitem[Schubart(1970)]{S70} Schubart, J.\ 1970, \iaucirc, 2268, 1

\bibitem[Schubart(1974)]{S74} Schubart, J. 1974, \aap, 30, 289

\bibitem[Sitarski \& Todororovic-Juchniewicz(1995)]{ST95} Sitarski, G.~\&
Todororovic-Juchniewicz, B.\ 1995, Acta Astronomica, 45, 673

\bibitem[Standish \& Hellings(1989)]{SH89} Standish, E.~M.~\& Hellings,
R.~W.\ 1989, Icarus, 80, 326

\bibitem[Tedesco(1989)]{T89} Tedesco, E. F. 1989, in Asteroids II, eds. R.  P.
Binzel, T. Gehrels, and M. S. Matthews (University of Arizona Press, Tucson),
1090

\bibitem[Tedesco et al.(1983)]{T+83} Tedesco, E. F., Taylor, R. C., Drummond,
	J., Harwood, D., Nickoloff, I., Scaltriti, F., Schober, H. J., \&
	Zappala, V. 1983, Icarus, 54, 23

\bibitem[Tholen(1984)]{T84} Tholen, D. J. 1984, Ph.D. Thesis, University of
Arizona

\bibitem[Thomas et al.(1997)]{T+97} Thomas, P. C., Binzel, R. P., Gaffey, M.
	J., Zellner, B. H., Storrs, A. D., \& Wells, E. 1997, Icarus, 128, 88

\bibitem[Veverka et al.(1989)]{VHST89} Veverka, J., Helfenstein, P., Skypeck,
A., \& Thomas, P.\ 1989, Icarus, 78, 14

\bibitem[Viateau \& Rapaport(1998)]{VR98} Viateau, B.~\& Rapaport, M.\ 1998,
\aap, 334, 729

\bibitem[Wagener \& Caldwell(1988)]{WC88}Wagener, R., \& Caldwell, J. 1988, ESA
	SP-281, 1206

\end{thebibliography}
\end{document}